     \documentstyle[12pt,epsfig]{article}   
     \newlength{\dinwidth}                       
     \newlength{\dinmargin}                      
\def\lsim{\mathrel{\rlap{\lower4pt\hbox{\hskip1pt$\sim$}}
    \raise1pt\hbox{$<$}}}                
\def\gsim{\mathrel{\rlap{\lower4pt\hbox{\hskip1pt$\sim$}}
    \raise1pt\hbox{$>$}}}                
\def\be{\begin{equation}}
\def\ee{\end{equation}}
\def\bea{\begin{eqnarray}}
\def\eea{\end{eqnarray}}

\def\bit{\begin{itemize}}
\def\eit{\end{itemize}}

\def\h5{\hskip5mm}

\newcommand{\GeV}{\,\mbox{GeV}}

\begin{document}
\hfill{\parbox[t]{4cm}{PITHA 99/16 \\ hep-ph/9907280 }}
\vspace*{10mm}
\begin{center}  \begin{Large} \begin{bf}
Hadronization Corrections to Jet Cross Sections \\
in Deep-Inelastic Scattering \\
  \end{bf}  \end{Large}
  \vspace*{5mm}
  \begin{large}
M.~Wobisch$^a$, T.~Wengler$^b$\\ \vskip2mm
  \end{large}
\end{center}
$^a$III. Physikalisches Institut, RWTH Aachen,
D-52056 Aachen, Germany \\
$^b$Physikalisches Institut, Universit\"at Heidelberg, 
D-69120 Heidelberg Germany \\
\begin{quotation}
\noindent
{\bf Abstract:}
The size of non-perturbative corrections
to high $E_T$ jet production in deep-inelastic scattering
is reviewed.
Based on predictions from fragmentation models,
hadronization corrections for different jet definitions are 
compared and the model dependence as well as the dependence on 
model parameters is investigated.
To test whether these hadronization corrections can be applied to 
next-to-leading order (NLO) calculations, jet properties and
topologies in different parton cascade models  
are compared to those in NLO.
The size of the uncertainties in estimating the hadronization
corrections is compared to the uncertainties
of perturbative predictions.
It is shown that for the inclusive $k_\perp$ ordered jet clustering
algorithm the hadronization corrections are smallest 
and their uncertainties are of the same size as the uncertainties
of perturbative NLO predictions.
\end{quotation}
%
\section{Introduction}
Before the prediction of a perturbative QCD calculation
(``parton-level'' cross section) can be compared to 
a measured ``hadron-level'' jet cross section, the size of 
non-perturbative contributions (``hadronization corrections'')
has to be estimated.
Advanced techniques based on ``power corrections''~\cite{powercor} 
are presently only available for the mean values of event shape 
variables and predict very large hadronization corrections for
most of the HERA kinematic range, preventing
these observables to be used for stringent tests of perturbative QCD.
For such tests observables with small hadronization corrections 
are needed, for example the production rate of jets with high transverse 
energies\footnote{Throughout the whole paper ``transverse energy''
always refers to transverse energies in the Breit frame, where
``transverse'' means the direction perpendicular to the 
z-axis, which is given by the direction of the incoming 
proton and the exchanged virtual photon.}.
Predictions of hadronization corrections to these observables 
are presently only available in the form of phenomenological
fragmentation models such as the Lund string model 
(as implemented in JETSET~\cite{jetset}) and the 
HERWIG cluster fragmentation model~\cite{clusterfrag}.
These models are implemented in event generators that include
leading order matrix elements and a perturbative parton cascade 
which is matched to the hadronization model.

Based on these models, hadronization corrections are compared
for different jet definitions, including a new angular ordered
jet clustering algorithm (``Aachen algorithm'').
The model dependence and the dependence on model parameters
is investigated.
These model estimates are usually needed for comparisons
of perturbative QCD in next-to-leading order (NLO) to measured
data distributions. We therefore also discuss the compatibility of
jet topologies in parton cascade models and in NLO.
Finally we review the uncertainties of NLO predictions
and compare their size to the uncertainties of the estimates
of hadronization corrections.

\section{Definitions}
The present study includes four different jet clustering algorithms
which differ in two aspects in how they define jets.
The first aspect is the {\bf ordering} in the clustering of particles.
This is either done in the order of smallest relative transverse 
momenta (``$k_\perp$ ordering'') or in the order of smallest angles
(``angular ordering'') between particles.
The second aspect concerns the definition of the jets inside the event.
In one case {\bf all} particles are clustered either to one of the 
hard jets or to the proton remnant (``exclusive'' definitions),
while in the other case only {\bf some} particles are clustered
into the hard jets, while other particles remain outside the hard 
jets (``inclusive'' definitions). 
The following four jet definitions are used, 
all in the Breit frame:
\begin{itemize} 
\item {\bf \boldmath the exclusive $k_\perp$ ordered algorithm} 
as proposed in \cite{exclkt}.

\item {\bf the Cambridge algorithm} as proposed in \cite{cambridge}
but modified for DIS to consider the proton remnant as a particle 
of infinite momentum according to the prescription in \cite{exclkt}. 
This algorithm is similar to the {\bf exclusive} $k_\perp$ but uses 
{\bf angular ordering}. 

\item {\bf \boldmath the inclusive $k_\perp$ ordered algorithm} as 
proposed in \cite{inclkt}.

\item {\bf the Aachen algorithm} --- this is a new jet definition, 
invented for these comparisons. In analogy to the modification from the 
exclusive $k_\perp$ algorithm to the Cambridge algorithm, we have 
modified the inclusive $k_\perp$ algorithm to obtain an {\bf inclusive} 
algorithm with {\bf angular ordering}.
The definition is very simple: particles with smallest
$R^2_{ij} = \Delta \eta^2_{ij} + \Delta \phi^2_{ij}$ are successively
clustered into jets, until all distances $R_{ij}$ between jets are 
above some value $R_0$ 
(as for the inclusive $k_\perp$ algorithm we set $R_0 = 1$).
The jets with highest $E_T$ are considered in the analysis. 
In dijet production in the Breit frame this definition is, at NLO, 
identical to the inclusive $k_\perp$ algorithm. 

\end{itemize}

For the exclusive jet definitions the recombination of particles 
is performed in the ``$E$-scheme'' (addition of four-vectors), 
while for the inclusive definitions it is done in the
``$E_T$-scheme'' (the $E_T$ of the jet is 
the scalar sum of the particle $E_T$s)~\cite{etscheme}.
To obtain jet cross sections of similar size for all jet definitions
the following parameters are used\footnote{These parameters are identical
to those used in a recent dijet analysis by the 
H1 collaboration~\cite{ichep520}.}:
\begin{itemize} 

\item for the inclusive jet definitions jets are required to have
$$
E_{T{\rm jet}} > 5\GeV \, , \hskip9mm  
E_{T{\rm jet1}} + E_{T{\rm jet2}} > 17\GeV
$$

\item for the exclusive jet definitions, the resolution of jets
in the event is defined by the resolution parameter $y_{\rm cut}$
$$
 y_{\rm cut} < k^2_{\perp ij} / 100\GeV^2  
\hskip9mm \mbox{with} \hskip2mm 
k^2_{\perp ij} = 2\,\min(E^2_i,E^2_j) \; (1- \cos \theta_{ij})
\hskip9mm \mbox{and} \hskip2mm 
y_{\rm cut} = 1\, .  
$$

\end{itemize}
Only events with (at least) two jets in the central
region of the detector acceptance ($-1 < \eta_{\rm jet, lab} < 2.5$)
are accepted.  
The studies are performed in the kinematic range $0.2 < y<0.6$
and $150 < Q^2 < 5000\GeV^2$ (unless stated otherwise).

We define the hadronization corrections to an observable ${\cal O}$ 
as the ratio of its value in a perturbative calculation
(``parton-level'': ${\cal O}_{\rm parton}$) and its value in a 
calculation including perturbative and non-perturbative 
contributions (``hadron-level'': ${\cal O}_{\rm hadron}$)
$
c_{\rm hadr. corr.} = {\cal O}_{\rm parton} \, / \, {\cal O}_{\rm hadron} \, .
$

All predictions have been obtained by the QCD models
HERWIG5.9~\cite{herwig} (using leading order matrix elements (LO ME), parton
shower and cluster fragmentation),  
LEPTO6.5~\cite{lepto} (LO ME, parton shower and string fragmentation) and 
ARIADNE4.08~\cite{ariadne} (LO ME, dipole cascade and string fragmentation).
The calculations have been performed for the HERA data-taking 
in 1997 ($820\GeV$ protons collided with $27.5\GeV$ positrons) using CTEQ4L 
parton distributions and the 1-loop formula for the running of $\alpha_s$.
The LEPTO predictions are obtained without the soft color 
interaction model. 
The NLO calculations are performed using the program DISENT~\cite{disent}
in the $\overline{{\rm MS}}$-scheme for CTEQ4M parton distributions 
and the 2-loop formula for the running of $\alpha_s$.
The renormalization scale is set to the average transverse energy
of the dijet system $\mu_r = \overline{E}_T$, the 
factorization scale to the mean $E_T$ of the jets
$\mu_f = \langle E_T \rangle \simeq 14\,{\rm GeV}$.

\section{Size and Model Dependence of the Predictions }

\begin{figure}
\centering
 \epsfig{file=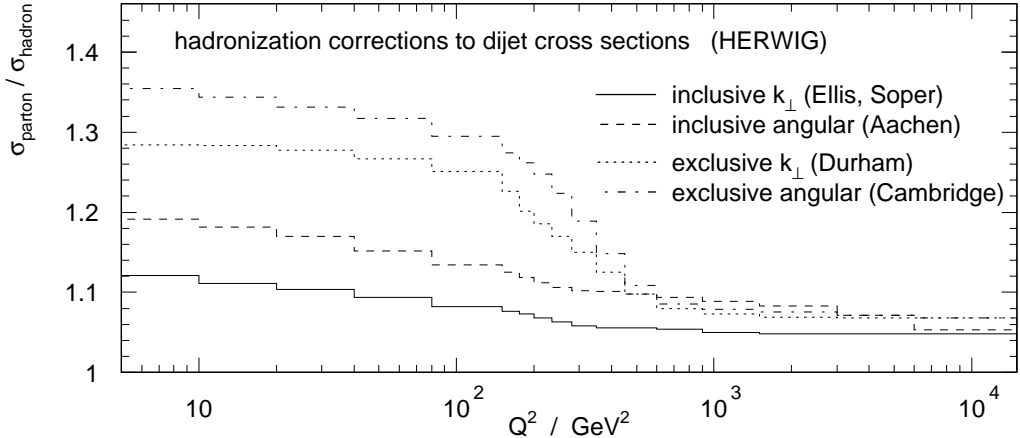}
\vskip-6mm
\caption{{\it The hadronization corrections to the dijet cross section 
for different jet definitions as a function of $Q^2$ 
as predicted by the HERWIG cluster fragmentation model.}}
\label{fig:hadcoralgo}
\end{figure}

\begin{figure}
\centering
 \epsfig{file=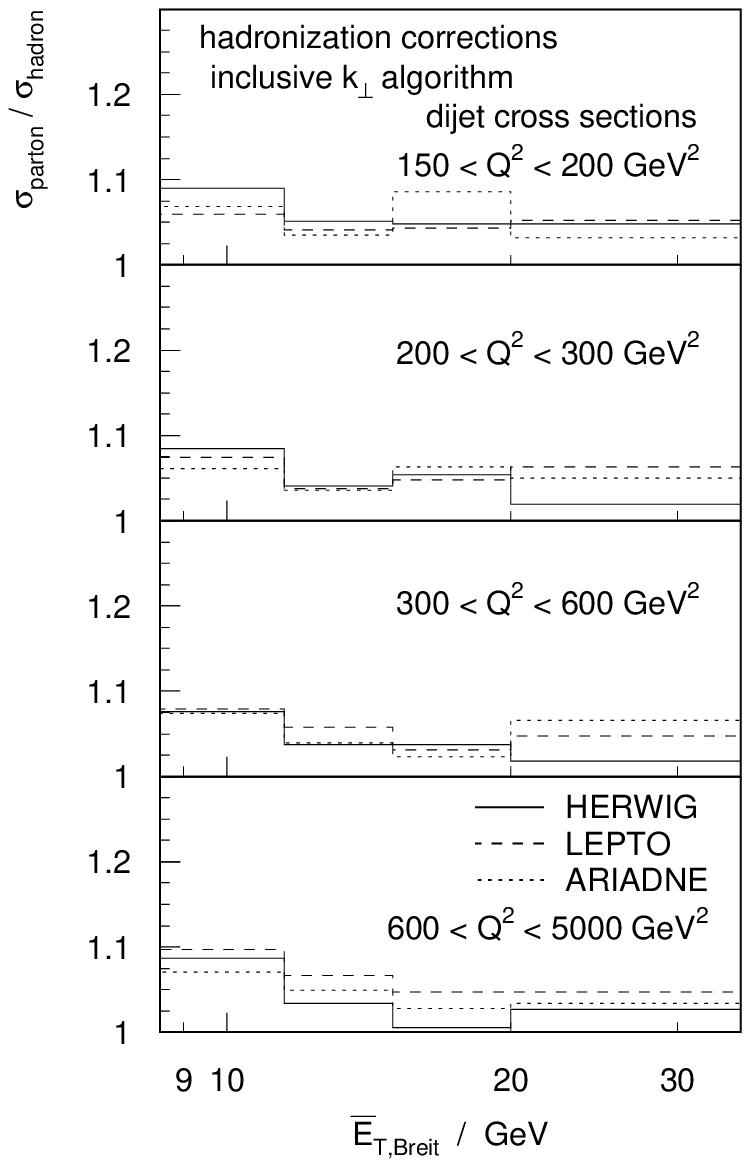,height=11.6cm,width=8cm}
 \epsfig{file=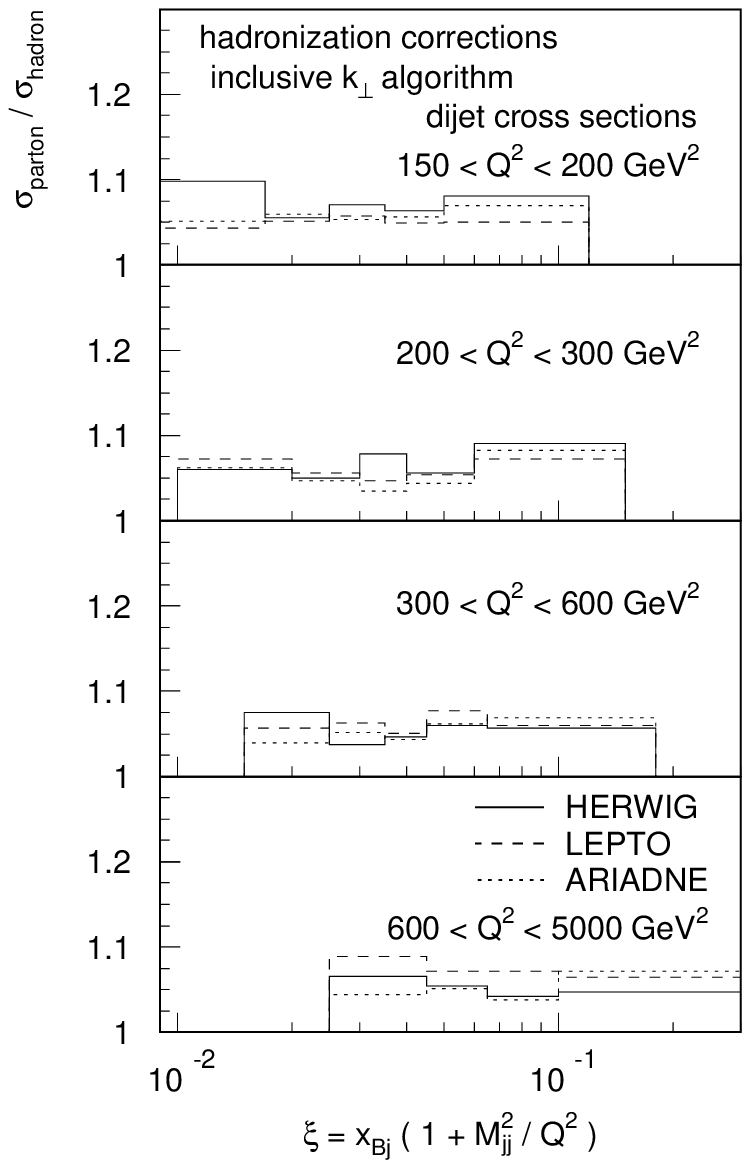,height=11.6cm,width=8cm}
\vskip-4mm
\caption{{\it Hadronization corrections for differential dijet distributions
for the inclusive $k_\perp$ algorithm as predicted from different
models.}}
\label{fig:hadcormodel}
\end{figure}

\begin{table}[b!]
\footnotesize
\centering
\begin{tabular}{l|r | r}
\hline
LEPTO / JETSET model parameters  & default & variation \\
\hline
\hline
$\Lambda_{QCD}$ in initial state parton shower 
     & 0.25\,\GeV & 0.25 -- 0.4\,\GeV \\
$\Lambda_{QCD}$ in final state parton shower 
      & 0.23\,\GeV & 0.23 -- 0.4\,\GeV \\
$Q_0^{ISR}$ cutoff for initial state parton shower & 1\,\GeV & 0.7 -- 2.0 \GeV\\
$Q_0^{FSR}$ cutoff for final state parton shower & 1\,\GeV & 0.5 -- 4.0 \GeV \\
width of Gaussian primordial $k_t$ of partons in the proton 
 & 0.44 \GeV& 0.44-- 0.7 \GeV\\
width of Gaussian distribution in $k_t$ when a non-trivial  & & \\
\hskip1cm target remnant is split into a particle and a jet 
& 0.35 \GeV& 0.35-- 0.7 \GeV\\
Gaussian width of $p_t$ for primary hadrons & 0.36 \GeV& 0.25 -- 0.45 \GeV\\
$a$ parameter in the symm.\  Lund fragmentation function & 0.3  & 0.1 -- 1.0 \\
$b$ parameter in the symm.\  Lund fragmentation function & 0.58 & 0.44 -- 0.7 \\
\hline
\hline
\end{tabular}
\label{table:parameters}
\caption{{\it Overview on the LEPTO and JETSET parameters and the 
ranges in which they are varied for the studies of the 
uncertainties of hadronization corrections.}}
\end{table}

The hadronization corrections as defined above
are shown in Fig.~\ref{fig:hadcoralgo} for the HERWIG model
as a function of $Q^2$ for the different jet definitions.
While at $Q^2 > 1000\GeV^2$ all jet definitions have similar and
reasonably small corrections (below 10\%), at smaller $Q^2$
large differences are seen.
In all cases the corrections are smaller for inclusive jet definitions 
than for exclusive definitions, and smaller for $k_\perp$ ordered
algorithms than for angular ordered ones.
Only the inclusive $k_\perp$ algorithm shows a small $Q^2$ dependence
and acceptably small corrections, even down to very small $Q^2$ values
(below 10\%).
For this definition we will study in more detail differential
distributions.
In Fig.~\ref{fig:hadcormodel} the hadronization corrections
from different models are shown as a function of the average 
transverse jet energy $\overline{E}_T$ and the reconstructed 
parton momentum fraction $\xi$ in different regions of $Q^2$.
While the corrections for the $\xi$ distribution are flat in all
$Q^2$ regions, we observe a slight decrease towards higher 
$\overline{E}_T$.
The predicted corrections agree within 3\% between the different models.

\begin{figure}
\centering
\epsfig{file=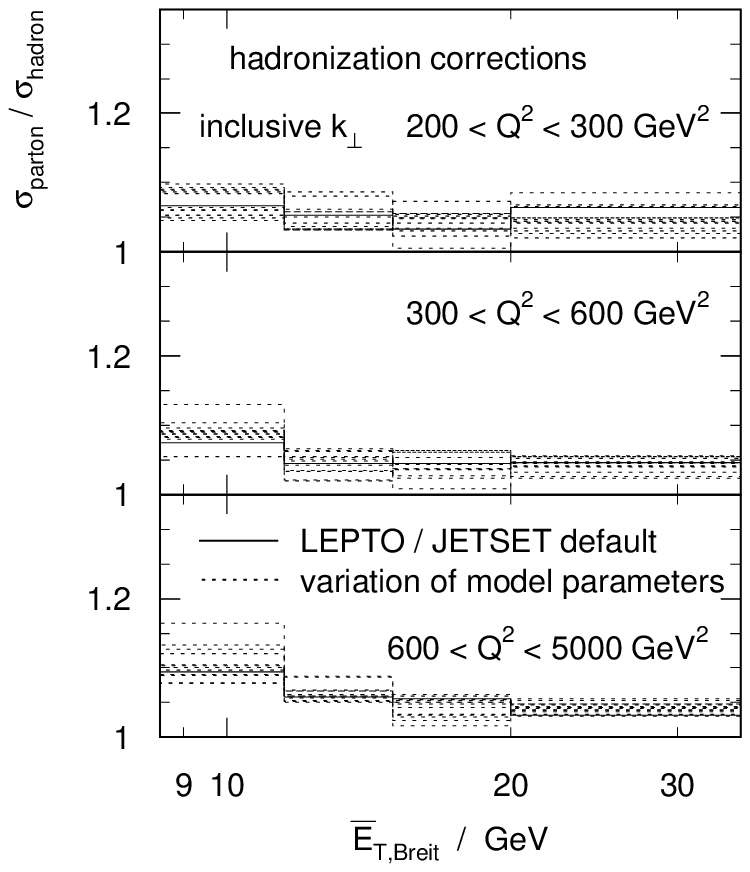}
\epsfig{file=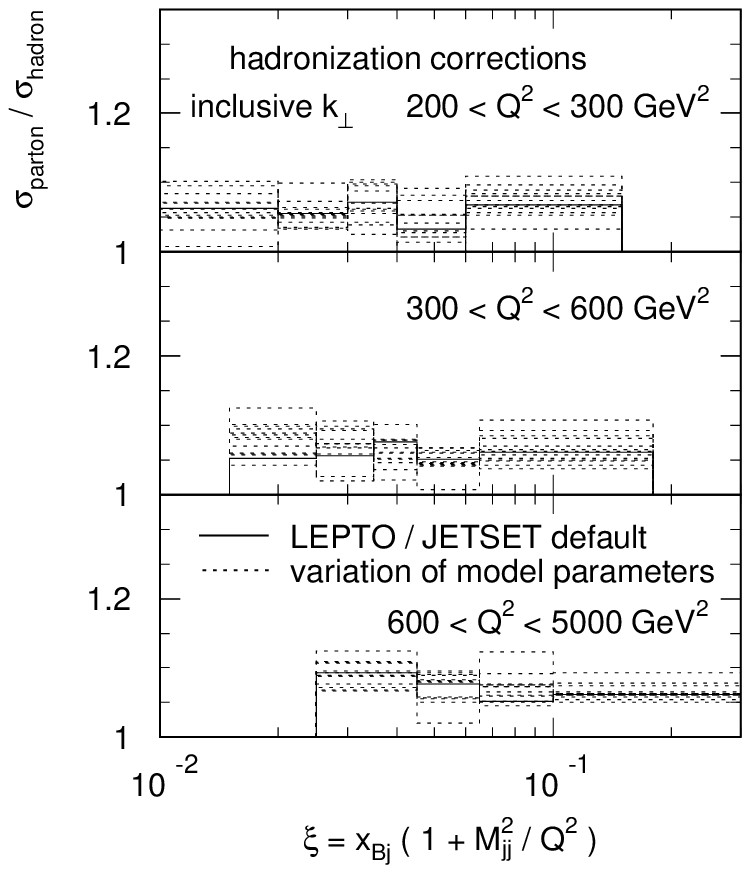}
\label{fig:parameters}
\caption{{\it The hadronization corrections to the 
$\overline{E}_T$ (left) and to the $\xi$ distributions (right) 
for the dijet cross section defined by the inclusive $k_\perp$ algorithm.
Shown are the predictions from the LEPTO/JETSET model with default
parameter settings (line) and the changes obtained by parameter variations
as described in the text (dotted lines). }}
\end{figure}

The predictions of these models may, of course, depend on 
parameters that define the perturbative parton cascade,
as well as on parameters of the hadronization model.
We have investigated the sensitivity of the LEPTO/JETSET model 
predictions to variations of some parameters as listed in 
Table~\ref{table:parameters}.
Fig.~\ref{fig:parameters} gives an overview on the effects of 
these variations which are seen to be small in all cases
(less than 4\% level).

\section{Parton Cascade Models vs.\ NLO Calculations}

\begin{figure}
\centering  
\epsfig{file=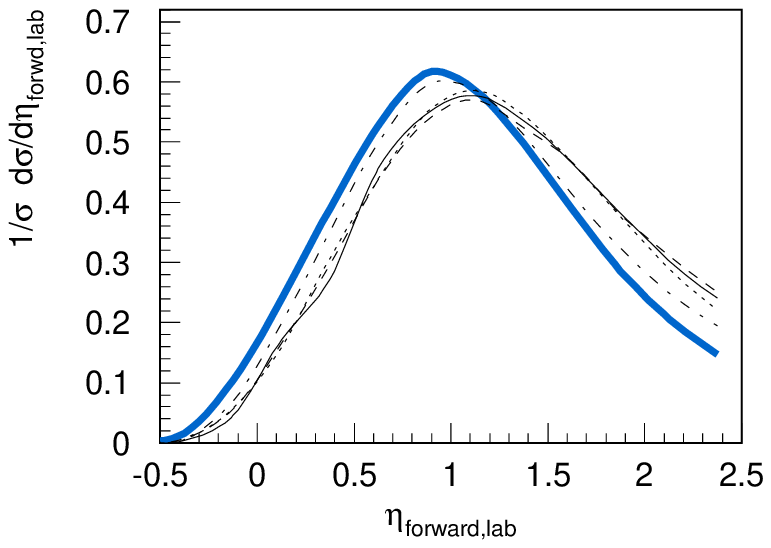}
\epsfig{file=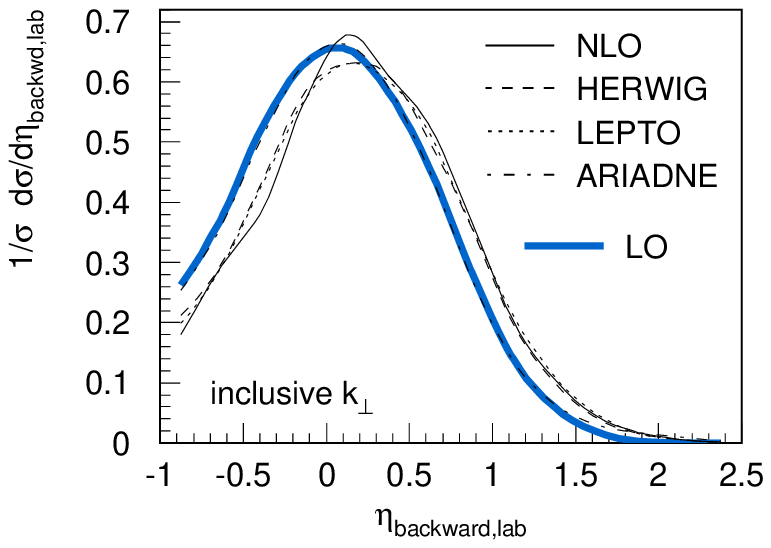}
\epsfig{file=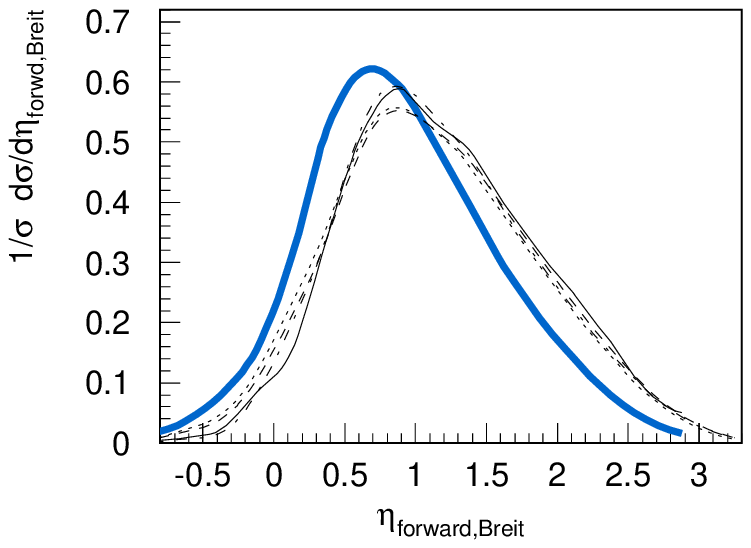}
\epsfig{file=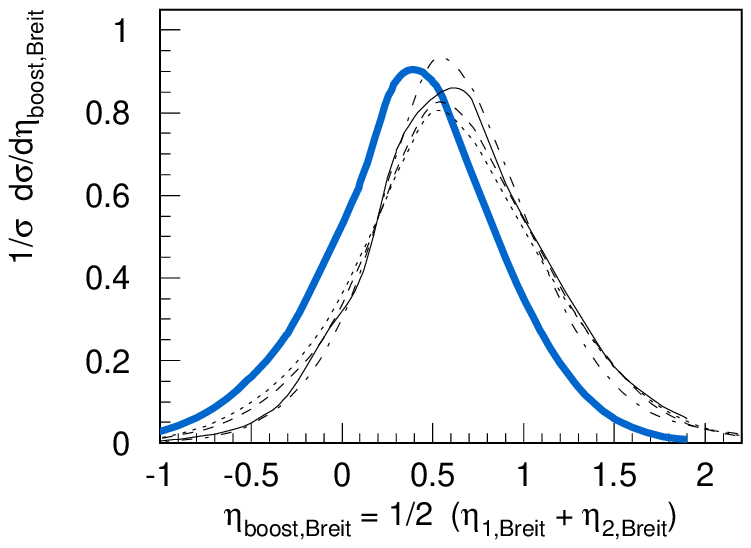}
\caption{{\it Higher order corrections to jet pseudorapidity  distributions
for the forward and the backward jet in the HERA laboratory frame, 
the forward jet in the Breit frame and the average pseudorapidity 
of the dijet system.
Displayed are the predictions from the leading-order matrix elements (LO)
and those including higher order corrections from either 
the next-to-leading order (NLO) calculation, or as given by
parton showers (HERWIG, LEPTO) or the 
dipole cascade (ARIADNE) for the inclusive $k_\perp$ algorithm.
Positive pseudorapidities are towards the proton direction in both 
the laboratory and the Breit frame.}}
\label{fig:jetrap}
\end{figure}

There is no unique way to separate perturbative and 
non-perturbative contributions in theoretical calculations.
A consistent treatment requires a well defined matching of 
both contributions, e.g. by the introduction of an
``infrared matching scale''~\cite{powercor}.
This, however, is not (yet) available for high $E_T$
jet cross sections in DIS.
The only available predictions are those of the hadronization models 
mentioned above.
So the following question appears:
``What are the uncertainties if we nevertheless use these model
predictions for estimating the hadronization corrections to be applied
to NLO calculations?''

Our attempt to tackle this problem is to assume that 
non-perturbative effects alter the production rates of 
multi-jet events only due to the change of the final state topology.
Hadronization effects for example cause particles to migrate out 
of the phase space considered for a particular jet, 
leading to a decrease of the jet's  transverse energy.
For a fixed $E_{T {\rm jet}}$ selection cut the resulting 
jet cross section will be reduced in this case.
The argument is therefore the following: If the final states in the NLO
calculation and parton shower models show the same properties, the
same influence of hadronization processes is to be expected for
both. In this case the model predictions can be used to estimate the
hadronization corrections for the NLO calculation.
In the following we address this question by comparing the 
predictions of the parton cascade models and NLO for the
distribution of jets inside the event (angular jet distributions),
the internal structure of the single jets (subjet multiplicities)
and the dependence of the dijet cross section on the $R_0$ 
parameter in the jet definition.

\subsection{Higher Order Corrections to Angular Jet Distributions}
The pseudorapidity distributions of jets are shown in 
Fig.~\ref{fig:jetrap} for the forward and the backward jet in the 
HERA laboratory frame (top), for the forward jet in the 
Breit frame (bottom left) and for the average jet pseudorapidity
of the dijet system (bottom right).

Compared are the predictions from the leading-order matrix elements
(LO, necessarily the same for DISENT, HERWIG, LEPTO and ARIADNE)
and those including higher order corrections from either 
the next-to-leading order, or as given by 
parton showers (HERWIG, LEPTO) or the 
dipole cascade (ARIADNE) for the inclusive $k_\perp$ algorithm.
All angular jet distributions are shifted by
the NLO corrections to the forward (i.e.\  the proton)
direction --- a feature
which is reproduced by all parton cascade models
(with the exception of the shift in $\eta_{\rm forward, lab}$ by ARIADNE).
We therefore do not derive any uncertainty on the estimation of
hadronization corrections for NLO from the study of the angular jet
distributions.

\subsection{Internal Jet Structure}
\begin{figure}
\centering
\epsfig{file=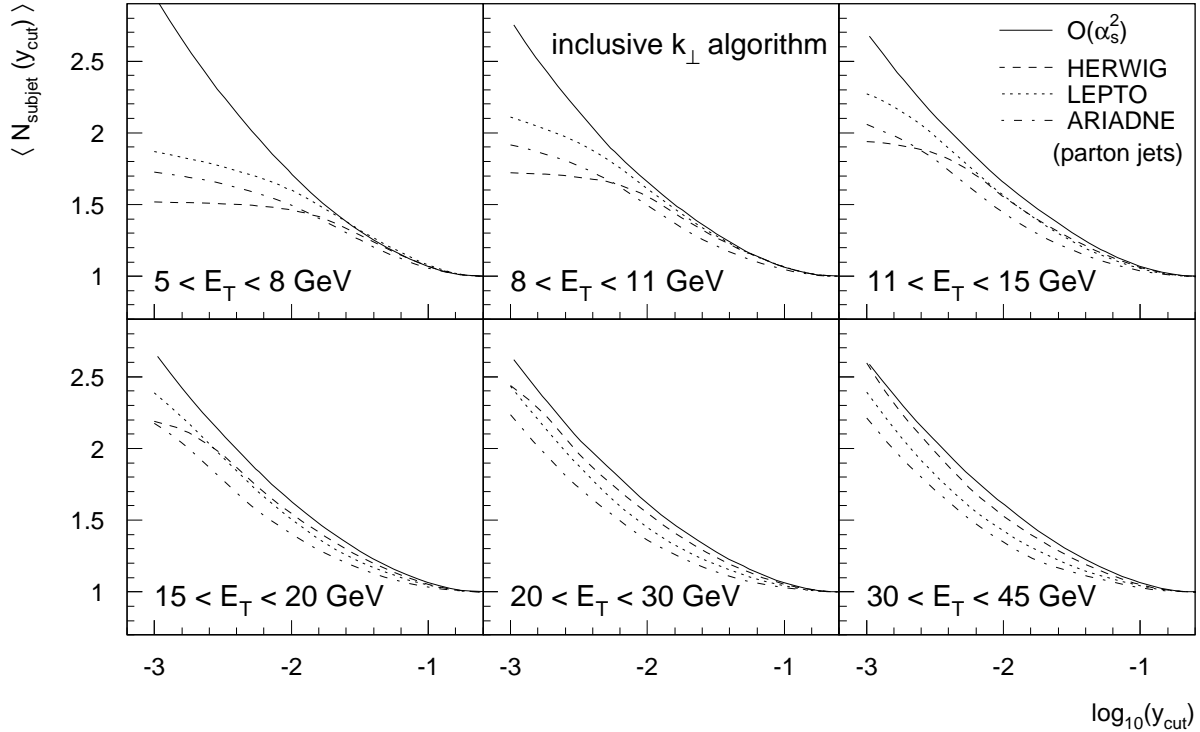}
\caption{{\it Subjet multiplicities for an inclusive jet sample
defined by the inclusive $k_\perp$ algorithm.
Compared are the predictions of different parton cascade models 
to the ${\cal O}(\alpha_s^2)$ calculation.}}
\label{fig:subjets} 
\end{figure}

Another test of the comparability of the different approaches
is the internal structure of jets. 
We decided to compare the average number of subjets that are
resolved at a resolution scale $y_{\rm cut}$ which is a fraction
of the transverse jet energy (a detailed definition of this observable
can be found in~\cite{h1jetstr}).
The average number of subjets is shown in Fig.~\ref{fig:subjets} 
as a function of the 
resolution parameter $y_{\rm cut}$ in different regions of  
$E_{T {\rm jet}}$ for an inclusive jet sample
in the same $\eta_{\rm lab}$ region where the dijet sample is defined.
These subjet multiplicities are sensitive to perturbative 
processes at larger $y_{\rm cut}$ values, while towards 
smaller $y_{\rm cut}$ non-perturbative contributions become 
increasingly important.

At smaller $y_{\rm cut}$ the ${\cal O}(\alpha_s^2)$ 
calculation\footnote{While the ${\cal O}(\alpha_s^2)$ calculation 
makes next-to-leading order predictions for dijet cross sections, 
it describes the internal structure of jets only at leading order.} 
has a very different behavior than the parton cascade models
where the number of subjets is limited by the available
number of partons due to the cutoff in the parton shower
while the ${\cal O}(\alpha_s^2)$ calculation
smoothly approaches the divergence at $y_{\rm cut} \rightarrow 0$.
These differences become smaller towards higher $E_T$ where 
both approaches show similar qualitative behavior, 
although significant differences remain.
Especially the dipole cascade in ARIADNE gives a much smaller
number of subjets than the ${\cal O}(\alpha_s^2)$ calculation.
The best agreement with the ${\cal O}(\alpha_s^2)$ 
calculation is observed for HERWIG at larger values of $y_{\rm cut}$
(which characterize the last steps in the clustering procedure).
Larger values of $y_{\rm cut}$ are hence connected to the coarse 
structure of jets which is the region where parton cascades
and NLO can be compared.

It is important to note that the spread of the models in
the relevant region of larger $y_{\rm cut}$ is of
the same order as their difference to the NLO calculation. In the
previous section we demonstrated that the predicted hadronization
corrections agree well between the different models.
We therefore conclude that 1) the hadronization corrections are not
sensitive to differences in the subjet multiplicities, and 2) that the
observed differences between model predictions and NLO in the subjet
multiplicities do not enter as an uncertainty in the estimation of the
hadronization corrections to NLO predictions.

\subsection{Radius Dependence of the Dijet Cross Section}
\begin{figure}
\centering
\epsfig{file=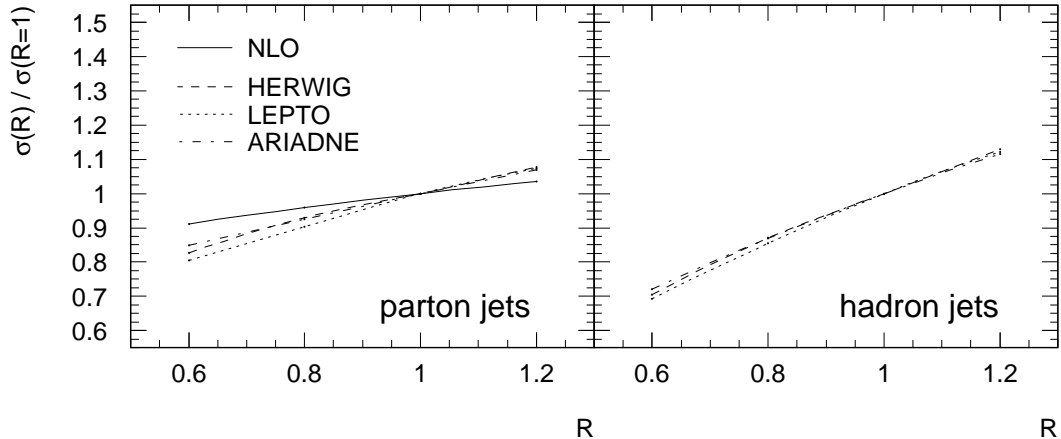,width=14.5cm}
\vskip-5mm
\caption{{\it The radius dependence of the dijet cross section for 
the inclusive $k_\perp$ algorithm.}}
\label{fig:raddep}
\end{figure}

The definition of the inclusive $k_\perp$ algorithm contains
a single free parameter $R_0$ which defines the maximal distance
within which particles are clustered in each step.
It follows that the final jets are all separated by distances
above $R_0$.

The dependence of the dijet cross section on the value of $R_0$
in the jet definition is directly correlated to the broadness
of the jets.
In Fig.~\ref{fig:raddep} we compare the $R_0$ dependence of the 
dijet cross section for parton jets (left) and for hadron jets (right).
Shown is the ratio of the dijet cross section at $R_0$
to the dijet cross section at $R_0 =1$ in the range $0.6 < R_0 < 1.2$.

For the (broader) hadron jets a large $R_0$ dependence
is observed (HERWIG: $\pm 13\%$ for a reasonable variation
$0.8<R_0<1.2$ around the prefered~\cite{inclkt} value of $R_0 =1$).
This dependence is reduced for the parton jets, but slightly
different for NLO ($\pm 4\%$) than for the parton cascade models
(HERWIG: $\pm 7\%$). 

The different behavior of the NLO dijet cross sections and the model
predictions for the same observable as a function of $R_0$ constitutes
an uncertainty when applying the model predictions of the
hadronization corrections to the NLO calculations. 
This difference (and the corresponding uncertainty) is however below 5\%.

\section{Properties of the Perturbative Cross Sections}
\begin{figure}
\centering
\epsfig{file=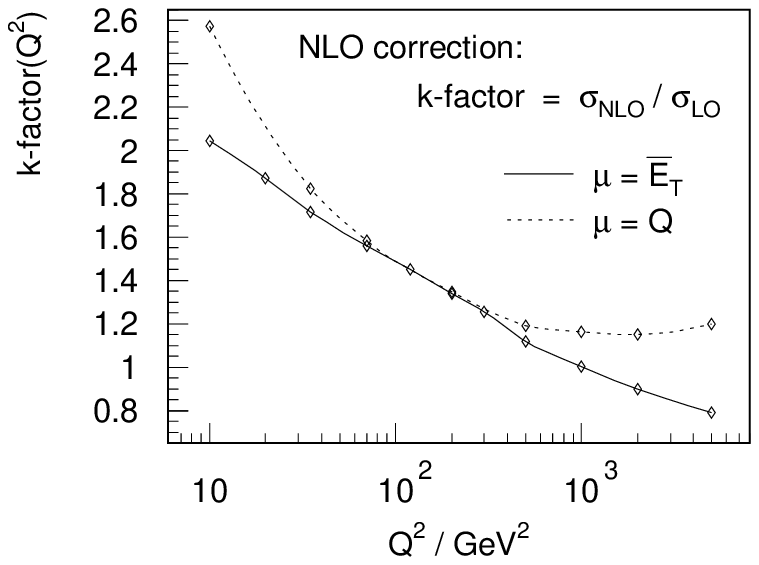,height=5.5cm,width=7.8cm}
\epsfig{file=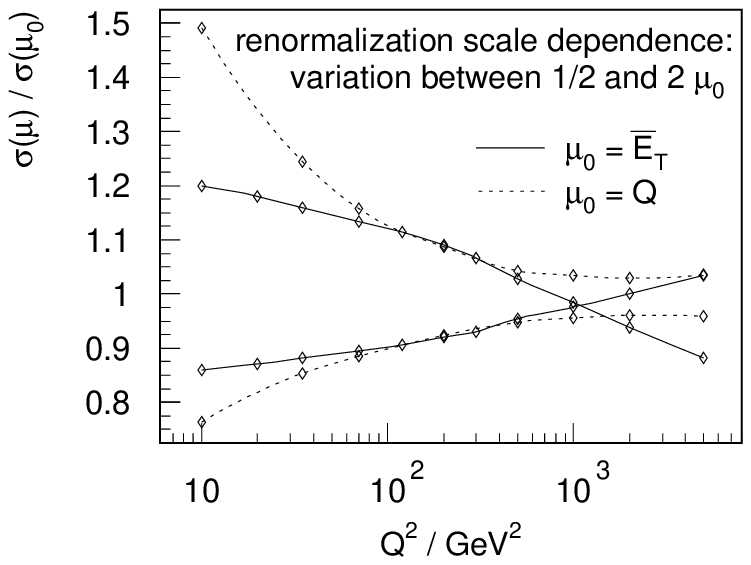,height=5.5cm,width=7.8cm}
\vskip-5mm
\caption{{\it The NLO correction to the dijet cross section 
    as a function of $Q^2$ for the
    inclusive $k_\perp$ algorithm for two different renormalization
    and factorization scales: $\overline{E}_T$ and $Q$ (left).
    The dependence of the NLO dijet cross section for the inclusive $k_\perp$
    algorithm on variations of the renormalization scale (right).}}
\label{fig:k+scale}
\end{figure}

In this section we give a brief overview on some properties
and uncertainties of the perturbative NLO cross sections.
Fig.~\ref{fig:k+scale} (left) shows the size of the NLO corrections
(i.e. the k-factor which is defined as the ratio of the NLO and the LO
cross section) for the inclusive $k_\perp$ algorithm as a function of $Q^2$.
To be sensitive to the fraction of the ${\cal O}({\alpha}_{s}^{2})$
contributions,
both the NLO and the LO calculations have been performed using 
the same (CTEQ4M) parton densities and the 2-loop formula for the 
running of $\alpha_s$.
The k-factor is shown for two different choices of the 
renormalization scale: $\mu_r = \overline{E}_T$ and $\mu_r = Q$.
For both scales the k-factor shows a strong dependence on $Q^2$.
While at large $Q^2$ the NLO corrections are small, they become 
sizeable for $Q^2 < 100\GeV^2$.
Throughout it is seen that the k-factor is smaller for a 
renormalization scale of the order of the transverse jet energies.

It is usually assumed that the scale dependence of a cross section
is somehow correlated to the possible size of higher order corrections,
and therefore a measure of the uncertainty.
Fig.~\ref{fig:k+scale} (right) shows the relative change of the dijet cross 
section when the renormalization scale $\mu^2$ is changed by a factor
of four up and down.
The comparison is made for the scales $\mu_r = \overline{E}_T$
and for $\mu_r = Q$.
The dependence on the 
renormalization scale becomes large at small $Q^2$.
Only for $Q^2 > 100\GeV^2$ this dependence is reasonably small
(below 10\%).
Over the whole range of $Q^2$ the renormalization scale dependence 
is smaller for the scale $\mu_r = \overline{E}_T$ than for the scale
$\mu_r = Q$.
The same variation has been studied for the factorization scale
and yields a negligible dependence (below 2~$\%$) over the whole range.

\section{Summary and Conclusions}
Hadronization corrections to jet cross sections in deep-inelastic 
scattering have been investigated based on predictions from 
the hadronization models HERWIG and JETSET as implemented in the event
generators HERWIG, LEPTO and ARIADNE.
It is seen that these corrections are smaller for inclusive and 
$k_\perp$ ordered jet definitions as compared to exclusive and angular 
ordered algorithms.
For reasonably large  transverse jet energies, the inclusive 
$k_\perp$ algorithm has hadronization corrections below 10\% over very
large regions of phase space.
The predictions  from different models are in very good agreement
and show only a weak dependence on the settings of specific
model parameters.
For the inclusive $k_\perp$ algorithm the corresponding uncertainties 
are not larger than 4\%.

A consequent and consistent consideration of hadronization corrections
for perturbative next-to-leading order (NLO) predictions requires 
a well-defined matching of a hadronization model to the NLO calculation. 
This is, however, not (yet) available.
Any other approach can only be an approximation and is subject to
various uncertainties.
Based on the assumption that these uncertainties are directly connected
to the differences in the final state topology in parton cascade 
models and in NLO calculations,
we have compared their predictions for various topological variables.
It is seen that changes in angular jet distributions w.r.t.\  leading
order calculations are very similar for the parton cascade 
models and NLO and hence do not contribute to the uncertainty. 
The subjet multiplicities show differences in their behavior in the NLO
calculations and the parton cascade models. These differences can
however be shown to have no significant influence on the 
predictions of the hadronization corrections. 
The dependence on the radius parameter $R_0$, which is directly 
correlated to the broadness of the jets, turns out to be different 
for NLO and the parton cascades, leading to an uncertainty
of less than 5\%.

We conclude that for the inclusive $k_\perp$ jet algorithm
at sufficiently large $Q^2$ and $E_{T {\rm jet}}$ the hadronization
corrections are under control with uncertainties not larger than 
those of the perturbative NLO calculations.
This will allow meaningful tests of perturbative QCD with a precision
of better than 10\%.



\begin{thebibliography}{99}
\bibitem{powercor}  M.~Dasgupta and B.R.~Webber,
 Eur. Phys. J. {\bf C1} (1998) 539.
\bibitem{jetset} T. Sj\"ostrand, Comp. Phys. Comm. 39 (1986) 347; \\
           T. Sj\"ostrand and M. Bengtsson, Comp. Phys. Comm. 43 (1987) 367. 
\bibitem{clusterfrag} B.R. Webber, Nucl. Phys. {\bf B238} (1984) 492
\bibitem{exclkt}  S. Catani, Yu.L. Dokshitzer and B.R. Webber,
Phys. Lett. {\bf B285} (1992) 291.
\bibitem{cambridge}  Yu.L. Dokshitzer, G.D. Leder, S. Moretti 
   and B.R. Webber, J. High Energy Phys. 08 (1997) 1.
\bibitem{inclkt} S.D.~Ellis, D.E.~Soper, Phys.~Rev. {\bf D48} (1993) 3160; 
    S. Catani, Yu.L. Dokshitzer,  M.H. Seymour and B.R. Webber, 
    Nucl. Phys. {\bf B406} (1993) 187.
\bibitem{etscheme} J. Huth et al., Proceedings of the Summer Study on 
          High Energy Physics, Snowmass, Colorado (1990) 134.
\bibitem{ichep520} H1 Collaboration, contributed paper 520 to ICHEP 98,
                    Vancouver, Canada (1998).
\bibitem{herwig} G. Marchesini et al., Comp. Phys. Comm. 67 (1992) 465.
\bibitem{lepto} G. Ingelman, A. Edin and J. Rathsman, Comp. Phys. Comm.
             101 (1997) 108.
\bibitem{ariadne} L. L\"onnblad, Comp. Phys. Comm. 71 (1992) 15.
\bibitem{disent}
S.~Catani and M.H.~Seymour, Nucl. Phys. {\bf B485} (1997) 291.
\bibitem{h1jetstr} H1 Collaboration, Nucl. Phys. {\bf B545} (1999) 3. 

\end{thebibliography}
\end{document}